\input{aipcheck}

\documentclass[
    ,final           
  ]
  {aipproc}

\layoutstyle{6x9}

\def\ltsima{$\; \buildrel < \over \sim \;$}
\def\ltsim{\lower.5ex\hbox{\ltsima}}
\def\gtsima{$\; \buildrel > \over \sim \;$}
\def\gtsim{\lower.5ex\hbox{\gtsima}}

\begin{document}

\title{The early X-ray afterglows of Gamma Ray Bursts}

\classification{98.70.Rz,95.85.Nv,96.60qe}
\keywords      {Gamma-ray burst, X-ray, Flares}

\author{G. Tagliaferri}{
  address={INAF - Osservatorio Astronomico di Brera, Via Bianchi 46, I-23807, Merate, Italy}
}

\begin{abstract}
The X-ray light curves of hundreds of bursts are now available, thanks to the X-ray Telescope
on board the Swift satellite, on time scales from $\sim 1$ minute up to weeks and in some 
cases months from the burst explosion. These data allow us to investigate
the physics of the highly relativistic fireball outflow and its interaction with
the circumburst environment. Here we review the main results of the XRT observations,
with particular regard to the evolution of the X-ray light curves in the early phases.
Unexpectedly, they are characterised by different slopes, with a very steep
decay in the first few hundred of seconds, followed by a flatter decay and, a few
thousand of seconds later, by a somewhat steeper decay. Often strong flare activity up
to few hours after the burst explosion is also seen. These flares, most likely, are still 
related to the central engine activity, that last much longer than expected
and it is still dominating the X-ray light curve well after the prompt phase, up to a 
few thousand of seconds. The real afterglow emission (external shock) is dominating the 
X-ray light curve only after the flatter phase ends. The flatter phase is probably the
combination of late-prompt emission and afterglow emission. When the late-prompt emission 
ends the light curve steepens again.
Some flare activity can still be detected during these later phases.
Finally, even the late evolution of the XRT light curves is puzzling, in particular many
of them do not show a ``jet-break''. There are various possibilities to explain
these observations (e.g.time evolution of the microphysical parameters, structured
jet). However, a clear understanding of the formation and evolution of the jet and
of the afterglow emission is still lacking. 
\end{abstract}

\maketitle


\section{Introduction}

The first afterglow associated to a Gamma Ray Burst (GRB) has been detected
in the X-ray band, thanks to {\it Beppo}SAX observations \cite{costa97}.
Optical and radio afterglows were soon discovered \cite{vapara97,frail97}.
These discoveries opened a new era in the studies of GRBs
and their associated afterglows. First of all, they showed that
GRBs are at cosmological distances and therefore they are the most powerful
explosions in the Universe after the Big Bang. Thanks to the firm association
of a GRB with a core-collapse SN in at least four cases, it is now generally 
believed that the progenitors of long-duration GRBs are massive stars, thus
supporting the collapsar model (see recent review
from \cite{woos06} and references therein).
While short-duration GRBs probably arise from
the merger of two compact objects; this is based on 1) their position inside 
their host galaxies (HG), 2) the properties of the HG and 3)
the properties of their light curves \cite{gehr05,covi06,fox05,barth05c,campa06}.

The studies in the pre-Swift era showed that the afterglows associated with
GRBs are rapidly fading sources, with X-ray and optical light curves characterised 
by a power law decay $\propto t^{-\alpha}$ with $\alpha \div 1-1.5$. Moreover, while
most of the GRBs, if not all, had an associated X-ray afterglow only about 60\% of them had
also an optical afterglow, i.e. a good fraction of them were dark--GRBs (see \cite{zame04}
for a general discussion on GRBs and their afterglows).
Therefore, it was clear that to properly study the GRBs,
and in particular the associated afterglows, we needed a fast-reaction
satellite capable of detecting GRBs and of performing immediate multiwavelength
follow-up observations, in particular in the X-ray and optical bands.
{\it Swift}  is designed specifically to study GRBs and their afterglows in multiple 
wavebands. It was successfully launched on 2004 November 20, opening
a new era in the study of GRBs \cite{gehr04}. {\it Swift} has on board three instruments:
a Burst Alert Telescope (BAT) that detects GRBs and determines their positions in the sky
with an accuracy better than 4 arcmin in the band 15-150 keV \cite{barth05a};
a X-Ray Telescope (XRT) that provides fast X-ray photometry and CCD spectroscopy
in the 0.2-10 keV band with a positional accuracy better than 5 arcsec \cite{burro05a};
an UV-Optical Telescope (UVOT) capable of multifilter photometry with a sensitivity
down to 24$^{th}$ magnitude in white light and a 0.5 arcsec positional accuracy
\cite{romi05}.

\begin{figure}
\includegraphics[height=.35\textheight]{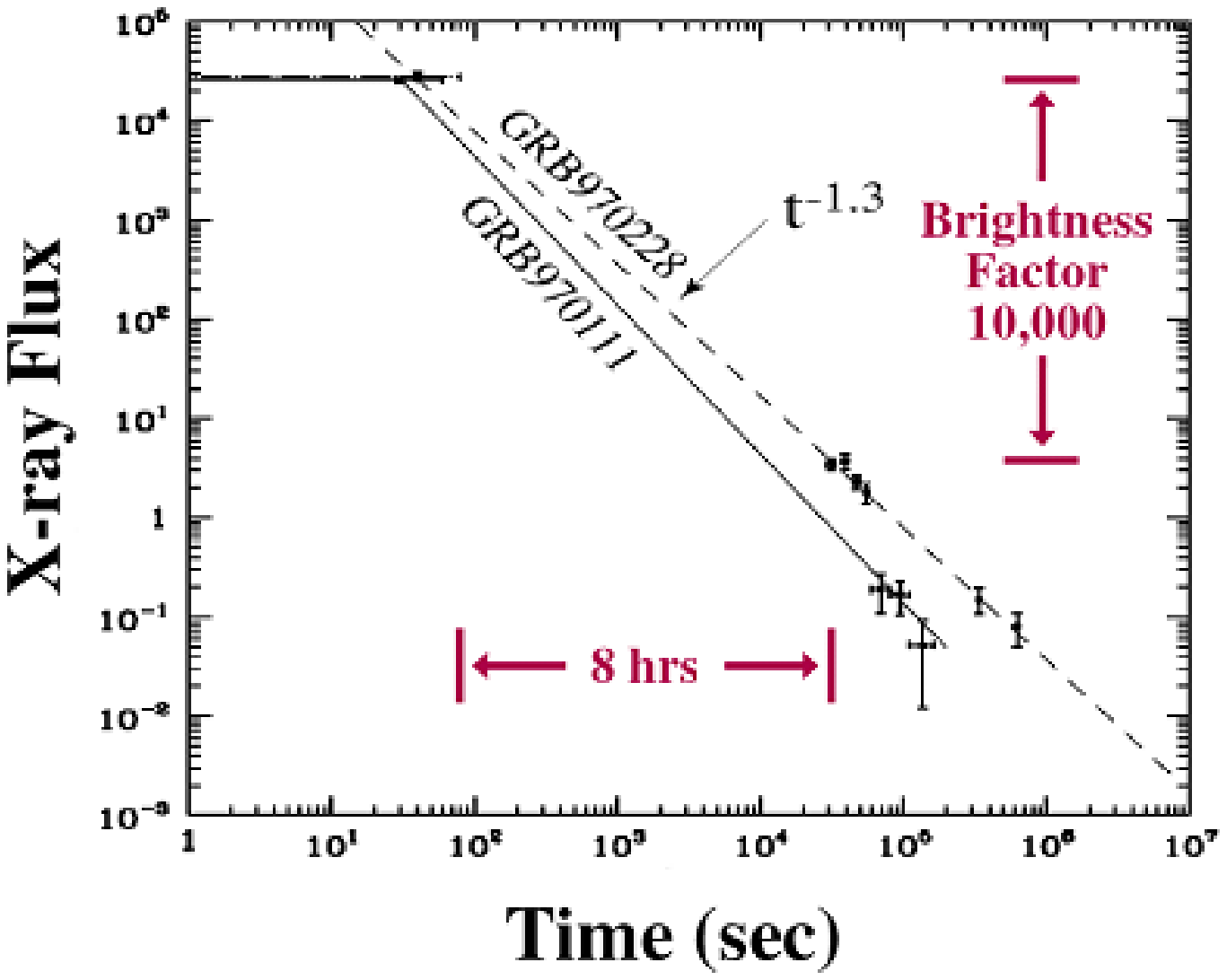}
\includegraphics[height=.34\textheight]{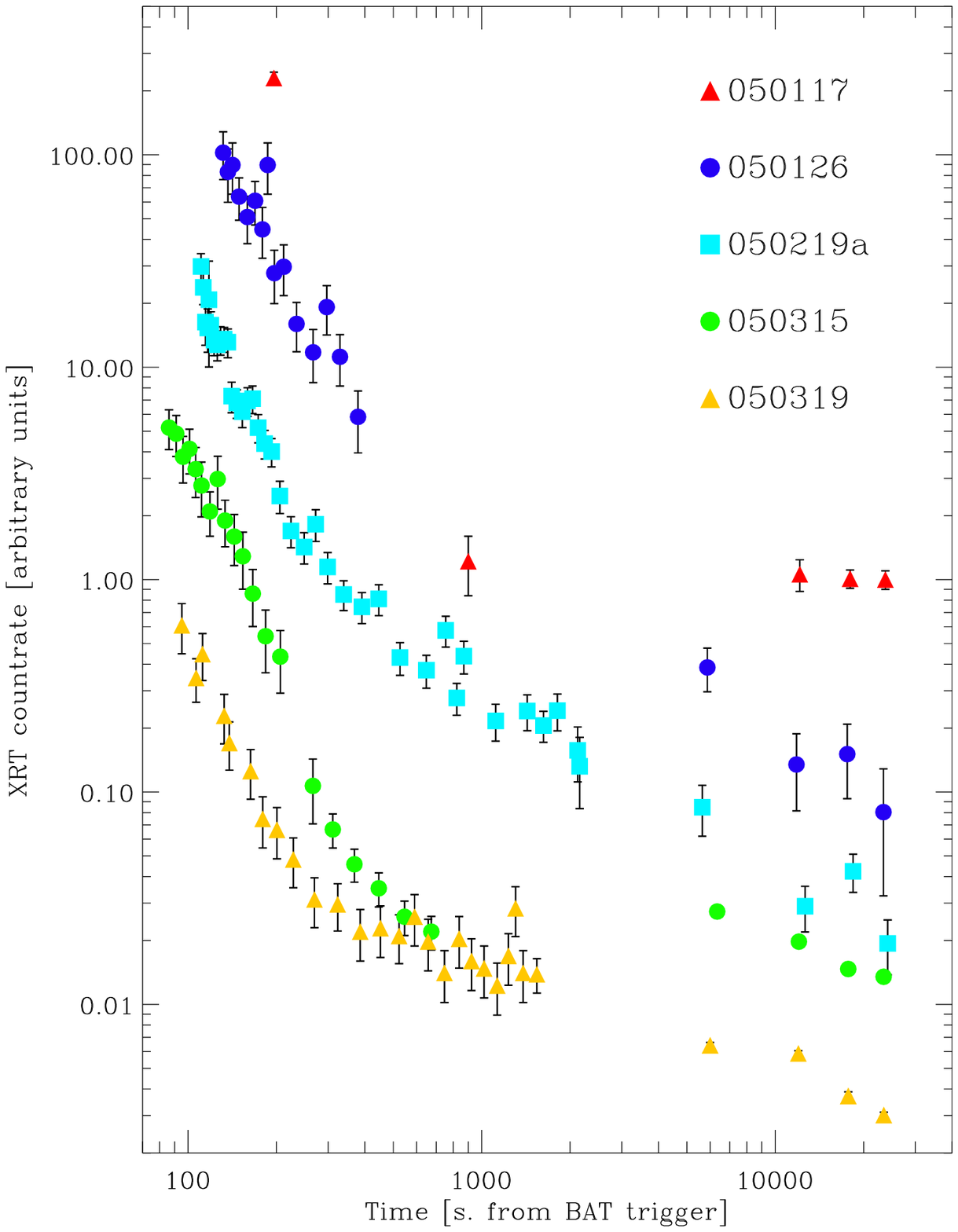}
\caption{{\it Left panel}: X-ray light curve of a typical GRB afterglows as
observed with {\it Beppo}SAX. Note how the backward extrapolations of the
afterglow light curves matched the flux of the burst itself. Therefore,
we were expecting a smooth power law decay of the X-ray afterglow light
curve since the first phases, gaining a few order of magnitude in source
brightness. {\it Right panel}: on the contrary a very steep decay and
then a flatteningis detected in the early
phases for the majority of the Swift GRBs (from \cite{taglia05}).}
\label{fg:lc_pw}
\end{figure}

In the first two years of operation Swift has detected about 200 GRBs.
Soon after detection the satellite autonomously determines
if it can repoint the narrow field instruments to the burst location
and, if possible, it usually slews to the source in less than 100-150 seconds.
Therefore, we have now X-ray light curves of hundreds of bursts that
cover a time interval from few tens of seconds up to weeks and months
for some of the bursts. As expected, the most spectacular results have
been obtained in the first few thousand seconds, i.e. in the gap not
covered by the previous missions. In particular, the XRT observations have
shown that the burst X-ray light curves in the early phases are much more 
complex than a simple backward extrapolation of the power law light curves 
observed few hours after the GRB explosion. Here we will outline the most
relevant results that have been obtained so far thanks to the XRT observations.

\section{The X-ray light curves}

\subsection{The early phases}

The X-ray observations obtained with {\it Beppo}SAX and other X-ray satellites
before the advent of Swift showed that the X-ray afterglow light curves from 
$>6$ hours after the explosion are well represented by a simple power law decay
with a decay index of the order of $\alpha \sim 1.4$. The backward extrapolation of
the afterglow X-ray flux matched that of the burst at the time of the explosion.
Therefore with the Swift satellite we were expecting to gain orders of magnitude in brightness
(see Fig.~\ref{fg:lc_pw}, left panel). Thanks to the much higher statistics we were then 
expecting to see, with higher signal to noise ratio (S/N), spectral lines that were
previously seen in the X-ray spectra of some afterglows, although with a not very 
high S/N (e.g. \cite{piro00,reeve03}).

\begin{figure}
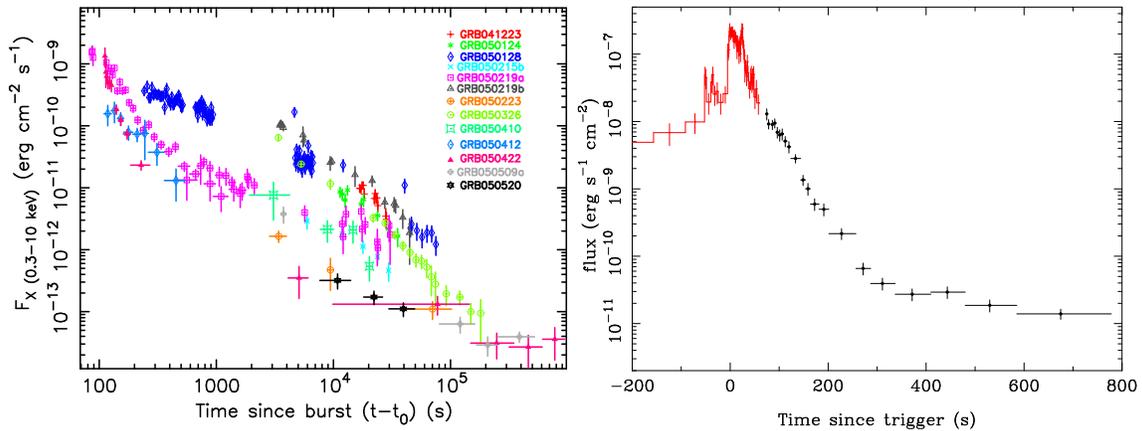

\includegraphics[height=.34\textheight,angle=270]{lc_xrt.ps}
\includegraphics[height=.34\textheight,angle=270]{grb050315.ps}
\caption{{\it Left panel}: the X-ray light curve of some Swift GRBs.
Note the different decaying behaviours detected in the early
phases and described in the text (from \cite{nouse06}).
{\it Right panel}: this figure shows the prompt BAT light curve converted
in the XRT X-ray band and the subsequent X-ray light curve as seen by XRT.
Note how the two light curves match perfectly, strongly supporting
the idea that the steep X-ray emission seen by XRT is an 
extension of the fading prompt emission (from \cite{vaugh06}).
}
\label{fg:lc_general}
\end{figure}

However, as often is the case when a new observing window became available,
the XRT data presented us with expected but also unexpected results. The XRT confirmed
that essentially all long GRBs are accompanied by a X-ray afterglow, there are only a couple
of them that have been fastly repointed by Swift and do not have an associated X-ray
afterglow (e.g. \cite{page06}). But, for instance, the XRT data do not show the presence of
spectral lines whatsoever in the X-ray spectra of GRB afterglows, neither in the first
few thousand second, nor at later (hours-days) time scales. They do show the presence
of a bright fading X-ray source. However, the source decay does not follow
a smooth power law, rather it is usually characterised by a very steep
early decay \cite{taglia05} (see Fig.~\ref{fg:lc_pw}, right panel), followed
by a flatter decay and then a somewhat steeper decay \cite{nouse06}
(see Fig.~\ref{fg:lc_general}, left panel).
Although this is the most common behaviour, in some of the Swift GRBs,
the early X-ray flux follows the expected and more gradual power law decay
(e.g. \cite{campa05,chinca05}). 

Do we have an explanation for what we are observing?
The most likely explanation for the steep early decay is that this is still due to the
prompt emission. Thanks to the fast reaction of the Swift satellite often we are
able to detect the prompt emission also with the XRT telescope and the steep
decay that we are observing is probably due to the ``high-latitude emission''
effect: when the prompt emission from the jet stops, we will still observe the
emission coming from the parts of the jet that are off the line of sight 
\cite{kuma00,taglia05,nouse06,obrie06,zhan06}. 
This interpretation is supported by the fact that the prompt BAT light curve
converted in the XRT band joins smoothly with that one seen by XRT for
almost all of the Swift GRBs \cite{barth05b,obrie06,vaugh06} (see Fig.~\ref{fg:lc_general}, right panel).
The origin of the flatter part that follow the early steep decay, that is well represented
by a power law with slope $0.5 \ltsim \alpha \ltsim 1$, is more controversial.
The total fluence that is emitted during this phase is comparable to, 
but it does not exceed that one of the prompt phase \cite{obrie06}.
It is probably a mixture of afterglow emission (the forward shock) plus a 
continuous energy injection from the central engine that refreshes the 
forward shock. When this energy injection stops, the light curve steepens
again to the usual power law decay already observed in the pre-Swift era
\cite{nouse06,ghise07}. 
Not all bursts show the steeper+flatter parts, a significant minority of 
them show a more gradual decay with $\alpha \ltsim 1.5$. These are more
consistent with the classical afterglow interpretation in which the
X-ray emission is simply due to the external shock. The flatter
part is not seen either because in these cases the continuous activity from
the internal engine is not present, or because the afterglow component is much 
brighter and it dominates over the internal contribution.

\subsection{The flares}

\begin{figure}
\includegraphics[height=.25\textheight]{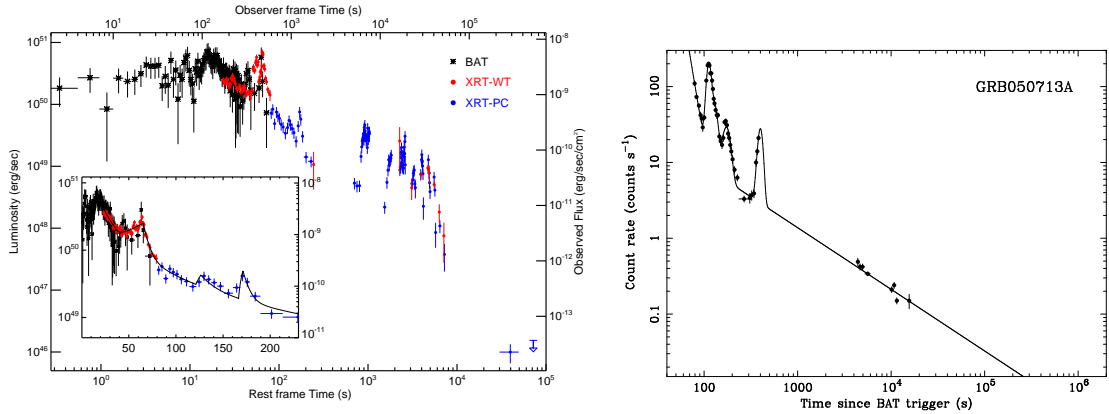}
\includegraphics[height=.25\textheight]{grb050713a.ps}
\caption{{\it Left panel}: the X-ray light curve of the very
high redshift, z=6.29, GRB050904, note the continuous flare
activity up to $10^4$ s in the source rest frame (from \cite{cusu07}).
{\it Right panel}: the X-ray light curve of GRB050713A, note the presence
of various strong flares both during the steep and flatter decay phases.
The underline X-ray light curve does not seem to be altered by these flares.}
\label{fg:flare}
\end{figure}

When XRT detected the first flares in the X-ray light curve of GRB050406
and then of GRB050502B \cite{burro05b,roma06,falco06}, this came as a
full surprise (although X-ray flares were already detected by {\it Beppo}SAX
in a couple of bursts, which were interpreted as due to the onset of
the afterglow \cite{piro05}). We now know that X-ray flares are present
in a good fraction of the XRT light curves (e.g. \cite{chinca07}).
Flares have been detected in all kinds of bursts: in X-ray flashes (XRF)
\cite{roma06}, in long GRBs (e.g. \cite{falco06,guetta06,paga06}), including
the most distant one at redshift 
z=6.29 (see Fig.~\ref{fg:flare}, left panel, \cite{cusu07}) and in short
GRBs \cite{barth05c,campa06}. These flares are usually found in the early 
phases up to a few thousand of seconds, but in some cases they are also
found at $> 10$ thousand seconds (see Fig.~\ref{fg:flare}). 
The ratio between their duration and peak time is very small, $\sim 0.1$, with
late flares having longer duration \cite{chinca07}. They can be very energetic
and in some cases can exceed the fluence of the prompt emission \cite{falco06}.
The fact that in the X-ray light curve of the same GRB there are more than one 
flare argues against the interpretation that the flares correspond to the onset
of the afterglow. Moreover, they do not seem to alter the underlying afterglow
light curve that after the flare follows the same power law decay as
before the flare (see Fig.~\ref{fg:flare}, right panel). Therefore, since
the beginning it was clear that these flares were correlated to the central
engine activities and not to the process responsible for the afterglow emission.

Also the spectral properties of these flares differ from those of the underlying
afterglow. For instance, their
broad-band spectral energy distribution is clearly formed by distinct
components. In particular, the optical-to-X-ray spectral index is often
much harder than both the optical and X-ray spectral indices alone. This
implies a spectral discontinuity between the two bands, again suggesting a
different origin for the two components \cite{taglia07}.
The X-ray spectra of the afterglows are well fitted by a simple power law
model plus absorption, with an energy spectral 
index of $\beta \simeq 1$. While the flares spectra are
usually harder and, for the strongest ones
that have better statistics, more complex models, such as a Band function
or a cutoff power law, are needed. Spectral evolution during flares is common, 
with the emission softening as the flare evolves, again a behaviour similar to
that seen during the prompt phase. Given this similarities between the prompt
and the flares properties, one would expect that X-ray flares are more common 
in those bursts with a prompt characterised by many pulses. But there seems
to be no correlation between the number of pulses detected in the prompt phase
and the number of X-ray flares detected by XRT. However, the distribution of the
intensity ratio of consecutive BAT prompt pulses and that one of consecutive XRT
flares is the same, another piece of evidence that prompt pulses and X-ray flares
have a common origin. For a comprehensive analysis of the flare properties
see references \cite{chinca07,falco07,liang06,koce07}.

Although various models have been proposed to explain the presence of these
X-ray flares, all these properties indicates that they are related to the
central engine activities and that they are due to the internal shocks,
rather than the external shocks \cite{chinca07}.

\subsection{The late X-ray light curve: any evidence for a jet break?}

In the standard fireball scenario (see \cite{mesza02} and references therein)
the afterglow emission is due to the deceleration of the expanding fireball
by the surrounding medium (external shock). If the expanding fireball is
collimated in a jet, then we expect to see an achromatic break in the power
law decay at the time when the full jet opening angle becomes visible to the
observer \cite{rhoad99}. The evaluation of the beaming factor is very
important in order to determine the total energy emitted by the burst,
in fact if we assume isotropic radiation this energy can range up to 
$10^{54} \ {\rm ergs}$. A value that is difficult to explain, unless
a beaming correction is applied. Breaks were detected a few days 
after the explosion in the optical and radio light curves of burst 
detected before the Swift advent. If interpreted as jet-breaks, then
the correct total energy emitted in the gamma band by the prompt 
clusters around $10^{51} \ {\rm ergs}$ \cite{frail01}. There seem to be
also a tight correlation between this energy and and the peak energy
of the prompt spectrum \cite{ghir04}. 

\begin{figure}
\includegraphics[height=.25\textheight]{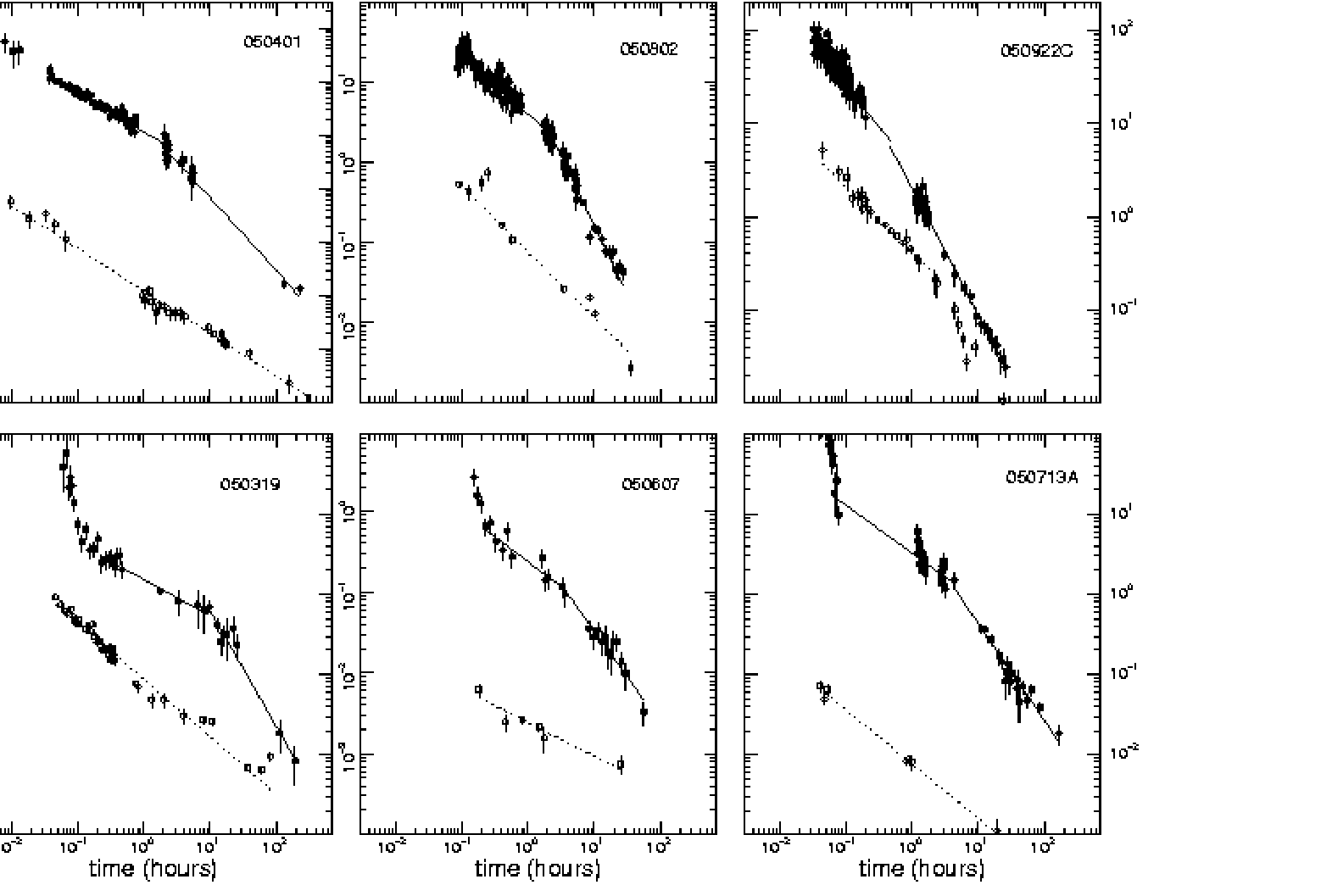}
\includegraphics[height=.25\textheight]{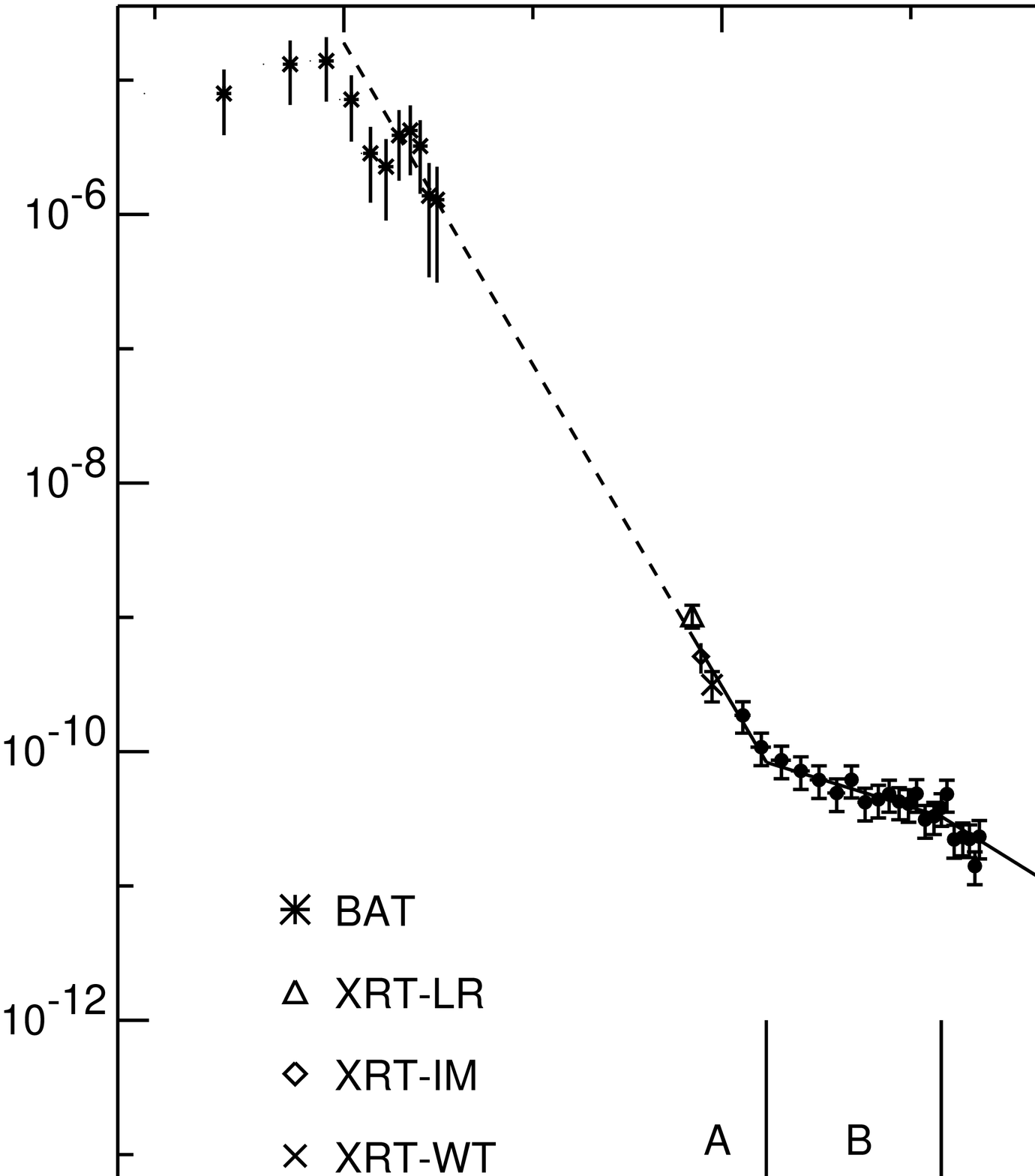}
\caption{{\it Left panel}: the X-ray and optical light curves of six Swift GRB
afterglows that show a chromatic X-ray break not seen in the optical
(from \cite{panai06}).
{\it Right panel}: BAT and XRT light curve of GRB050416A. The solid line represents
a double-broken power law model fit. Note the absence of any jet break up to
about 60 days after the burst. This absence is not consistent with the
empirical relations between the source rest-frame peak energy and 
the collimation-corrected energy of the burst
(from \cite{manga06}).
}
\label{fg:chromatic}
\end{figure}

If these breaks are really due to a jet, then they should be seen 
simultaneously also in the X-ray band. Before the advent of Swift the
observations in the X-ray band were limited and there were only few
measurements. Now thanks to XRT we have many detailed X-ray light curves
and the picture is not so clear any more. First of all as we have seen,
in the early phases there can be more than one break, but none of them
seems to be due to a jet-break. Rather they are probably due to the activity
of the internal engine, as we have seen previously. Moreover, for
some of these bursts we have also the early optical data and the breaks
are not seen in the optical, therefore they are not achromatic
(see Fig.~\ref{fg:chromatic}, left panel).
This behaviour can be explained either by assuming an evolution
of the microphysical parameters for the electron and magnetic energies
in the forward shock or by assuming that the X-ray and optical emission
arise from different components \cite{panai06}. 
In any case, from a systematic analysis of the XRT light curves of 107 GRBs,
72 afterglow breaks are found, but of these only 12 are consistent
with being jet-breaks and only 4 are not related to the early flat phase
\cite{willi07}.
In other words there are only 4 breaks that are good candidates for being 
jet breaks. Therefore, contrary to the earlier expectations, jet-breaks
seem to be the exception and not the rule in the X-ray light curves of
GRB afterglows (for a discussion on this argument see also \cite{burro07}). 
Moreover, by assuming the correlation between the prompt 
peak energy and the beaming corrected prompt energy derived for some GRB 
with a measured optical break \cite{ghir04}, we can check if the absence
of a jet-break in the X-ray light curve (see Fig.~\ref{fg:chromatic}, right panel)
is consistent with this correlation.
The result is that many of the XRT afterglows are outliers of this
correlation \cite{sato06,willi07} (however, note that the presence of these outliers
is still argument of discussion, see \cite{ghirla07}). 
If confirmed, this means that either this correlation
somehow is valid only for breaks observed in the optical and not in the X-ray
or that it is valid only for a subsample of GRBs whose properties have
still to be defined or that it is not as tight as previously thought.

\section{Conclusions}

After more than two years of Swift operations, the data provided by the XRT
allowed us to make break-through discoveries in various field of the GRB studies
including the detection of the afterglows of short GRBs. We did not discuss
this argument here, but for the first time we have been able to study in more details
the properties of these elusive sources and to find and study their host galaxies
with on ground follow-up \cite{gehr05,barth05c,campa06,fox05,laparo06,romi06,naka07,lee07}.
Thanks to the Swift fast repointing
and its instrumentation capabilities, we have now the fast localisation of GRB with 
an accuracy of few arcsec, which allows us to immediately start ground-based
observations. Uniform multiwavelength light curves of the afterglows are available
starting from $\sim 1$ minute after the burst trigger. In particular, in the X-ray
band, thanks to XRT, we have hundreds of light curves spanning the range from few tens 
of seconds up to weeks and months after the explosion. These data allow us to investigate
the physics of the highly relativistic fireball outflow and its interaction with
the circumburst environment.

Unexpectedly, these X-ray light curves are characterised by different slopes 
in the early phases and often by the presence of strong flare activity up to few 
hours after the burst explosion. The picture that is consolidating is that
the central engine activity lasts much longer than expected and it is still
dominating the X-ray light curve well after the prompt phase, up to a few
thousand of seconds. The external shock, the real afterglow, takes over
the emission only after the end of the flatter phase, although
some flare activity can be still detected during these later phases.
Finally, even the evolution of the XRT light curve at the later phases is providing
more questions than solutions. In particular, the lack of a ``jet-break'' in 
many of these light curves is puzzling. There are various possibilities to explain
these observations (e.g.time evolution of the microphysical parameters, structured
jet). However, a clear understanding of the formation and evolution of the jet and
of the afterglow emission is still lacking.


\begin{theacknowledgments}
This work was supported by ASI grant I/R/039/04 and MIUR grant 2005025417.
We gratefully acknowledge the contributions of dozens of members 
of the XRT and BAT teams at
OAB, PSU, UL, GSFC, ASDC, and MSSL and our subcontractors, 
who helped make these instruments possible.
\end{theacknowledgments}




\end{document}